\documentclass[aps,prl,twocolumn,groupedaddress,showpacs]{revtex4}
\usepackage{graphicx}
\usepackage{amsmath}

\newcommand{\ef}{E_{\scriptscriptstyle F}}
\newcommand{\kf}{k_{\scriptscriptstyle F}}

\newcommand{\leqa}{\stackrel{<}{\scriptstyle \sim}}

\newcommand{\rb}{\bar{\rho}}
\newcommand{\rt}{\widetilde{\rho}}
\newcommand{\Db}{\bar{\Delta}}
\newcommand{\Dt}{\widetilde{\Delta}}

\newcommand{\rms}[1]{\sqrt{\left< #1^2 \right>}}
\newcommand{\secmom}[1]{\left< #1^2 \right>}
\newcommand{\tm}{\tau_{\rm min}}
\newcommand{\tD}{\tau_{\scriptscriptstyle \Delta}}
\newcommand{\tH}{\tau_{\scriptscriptstyle H}}
\newcommand{\eps}{\varepsilon}

\begin{document}

%\title{Mesoscopic Fluctuations of the Pairing Gap}
%\title{Semiclassical Theory of the BCS Pairing Gap Fluctuations}
\title{Semiclassical Theory of Bardeen-Cooper-Schrieffer Pairing-Gap Fluctuations}
\author{H. Olofsson,$^1$ S. \AA berg,$^1$ and P. Leboeuf$~^2$}

\affiliation{$^1$Mathematical Physics, LTH, Lund
  University, P.O. Box 118, S-221 00 Lund, Sweden \\
$^2$Laboratoire de Physique Th{\'e}orique et Mod{\`e}les
Statistiques, CNRS, B{\^a}t. 100, Universit{\'e} de Paris-Sud, 91405
Orsay Cedex, France}

\begin{abstract}
Superfluidity and superconductivity are genuine many-body
manifestations of quantum coherence. For finite-size
systems the associated pairing gap fluctuates as a function of
size or shape. We provide a theoretical description of the zero temperature
pairing fluctuations in the weak-coupling BCS limit of
mesoscopic systems characterized by order/chaos dynamics. The
theory accurately describes experimental observations of nuclear
superfluidity (regular system), predicts universal fluctuations of
superconductivity in small chaotic metallic grains, and
provides a global analysis in ultracold Fermi gases.
%A description of quantum fluctuations of pairing properties in
%finite-sized quantum systems based on periodic orbit theory is
%presented. The size of the fluctuations are found to depend on
%quite general system properties. We distinguish between systems
%where corresponding classical motion is regular or chaotic ({\it
%or diffusive??}), and describe in detail fluctuations of the BCS
%gap as a function of the size of the system. We apply the obtained
%result for one regular system, pairing of nuclear ground states,
%and one chaotic system, superconductivity of small metallic
%grains.
\end{abstract}


\maketitle

A microscopic theory of superconductivity based on pairing was set
up in 1957 by Bardeen, Cooper and Schrieffer \cite{BCS}. These
theoretical ideas were subsequently applied to finite systems by
Bohr, Mottelson and Pines to describe ground-state superfluid
properties of atomic nuclei \cite{Pines}. Today pairing effects
are central in a broad range of quantum systems, including neutron
stars, metallic grains, atomic gases, nuclei, etc
\cite{RevNucl,NanoRev,RevFermiGas,Brink}.
As the system size diminishes, finite--size effects become
important and lead to corrections with respect to the bulk
homogeneous behavior. Of particular interest is the influence
of the discreteness of the
single--particle quantum energy levels. In connection with
superconductivity, its importance was initially emphasized by P.
W. Anderson \cite{Anderson}, who pointed out that
superconductivity in small metallic grains should disappear when
the single--particle mean level spacing becomes of the order of
the pairing gap. The validity of this criterion was qualitatively
confirmed experimentally in the 90's \cite{RBT}. Another
consequence of the discreteness of the energy levels is the
appearance of fluctuations as a parameter is varied. There is at
least one clear experimental evidence of fluctuations of the
pairing gap in superfluid systems, through the odd-even staggering
of nuclear masses as a function of the mass number.

%Mesoscopic quantum fluctuations appear in many different contexts,
%starting from shell effects in nuclear structure, to persistent
%currents or magnetization in mesoscopic condensed matter physics.
%Particularly relevant are recent experiments in ultracold Fermi
%gases, that offer new prospects for studying the superfluid phase
%transition, of which a direct observation was
%recently reported \cite{Zwierlein}.

\begin{figure}[t]
%\vspace{-1cm}
%\includegraphics[width=7cm,clip=true,angle=-90]{nuclpic_wsphere1.eps}
\includegraphics[width=7.3cm,clip=true]{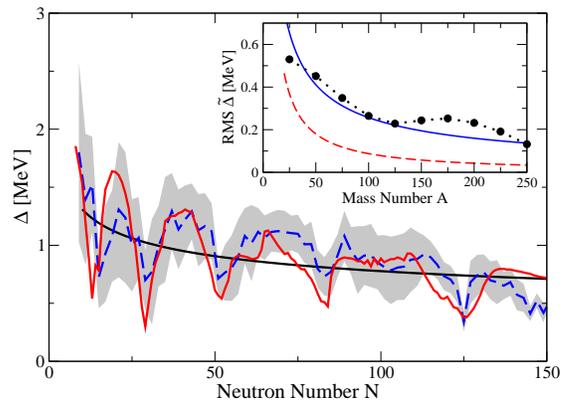}
\vspace{-0.3cm} \caption{Nuclear pairing gaps for neutrons. The
blue dashed line shows average experimental gaps, and isotopic
variation within one standard deviation is marked by the shaded
area. Calculations from the cavity model are shown by the solid
red line. Average behavior (Eq.~(\ref{Dbeq})) is shown by solid
black line. Inset: RMS of pairing gap fluctuations versus mass
number $A$. The dots are experimental data for protons and
neutrons. Solid and dashed lines are regular
(Eq.~(\ref{secmomreg})) and chaotic (Eq.~(\ref{secmomch}))
predictions, respectively. Data from Ref.~\cite{NuclMass}.}
  \label{Fignucl}
\end{figure}

Our purpose here is to present, within a mean--field
approximation, a theory of the pairing--gap fluctuations valid in
the weak coupling BCS limit for arbitrary ballistic potentials.
Our method is based on periodic orbit theory,
which has been successful in describing mesoscopic
fluctuations of thermodynamic and transport properties in
many--body systems \cite{semicl}.
%Periodic orbit theory is applied to a mean field description of
%the pairing field (in the abstract gauge space).
The results allow for a detailed calculation of
the fluctuations in specific systems. In particular, they provide
an accurate description of the pairing fluctuations in nuclei
(cf Fig.~\ref{Fignucl}). We also focus on statistical properties,
which are shown, generically, to be non-universal. The analysis
leads to a global and complete picture of the typical size of the
fluctuations in terms of properties of the corresponding classical
system, namely regular or chaotic dynamics.

%An attractive short-range residual interaction between fermions
%may give rise to a paired ground-state solution corresponding to a phase
%transition from the un-paired ground state.
%The attractive short-range residual interaction can
%be approximated by a seniority interaction, $V_{\rm pair}= -G
%P^{\dagger}P,$ where $P^{\dagger}$ creates a pair in time reversed
%orbits. In the mean field approximation, $G\left<
%P^{\dagger}\right> =\Delta$, the paired ground-state solution can
%be obtained in BCS theory \cite{BCS}. This implies solving the
%equation,
Our starting point is the mean field BCS equation for the pairing gap
$\Delta$ \cite{BCS},
\begin{equation} \label{gap1}
  \frac{2}{G}=\int_{-L}^L
  \frac{\rho(\eps)d\eps}{\sqrt{\eps^2+\Delta^2}},
\end{equation}
where $G$ fixes the strength of the pairing (seniority) interaction,
$\rho(\eps)$ is the single--particle level density, and we have put the
Fermi energy to zero. The energy cut off
$\pm L$ is given by the physical conditions, that are often
related to the determination of the force strength $G$.
Following semiclassical approaches, we divide the
pairing gap as well as the single-particle density of states
in a smooth part and a fluctuating part, $\Delta=\Db+\Dt$ and
$\rho=\rb+\rt$, respectively. In the weak coupling limit $\Delta \ll L$,
the smooth part of the gap is given by the well known solution
$\Db = 2L \exp (-1/\rb G)$ (see Ref.~\cite{PapenBulgacYu}
for regularization schemes). The fluctuating part of the density
$\rt$ can be expressed as \cite{semicl}
\begin{equation}
 \rt (\eps) = 2 \sum_p \sum_{r=1}^{\infty} A_{p,r} \cos(rS_p / \hbar + \nu_{p,r}),
\end{equation}
where the sum is over all primitive periodic orbits $p$ (and their
repetitions $r$) of the classical underlying effective
single-particle Hamiltonian. Each orbit is characterized by its
action $S_p$, stability amplitude $A_{p,r}$, period
$\tau_p=\partial S_p/\partial \eps$ and Maslov index $\nu_{p,r}$
(all evaluated at energy $\eps$). Assuming $\Dt \ll \Db$, an
equation for $\Dt$ may be obtained by multiplying Eq.~(\ref{gap1})
by $\Delta$, replacing the single--particle level density by its
semiclassical expression, and expanding up to lowest order in
fluctuating properties. Assuming moreover $\Db \ll L$ gives
\begin{equation} \label{dosc}
  \Dt = \frac{\Db}{\rb}\sum_p\sum_{r=1}^\infty A_{p,r}Y(r\tau_p)\cos \left(
  \frac{rS_p}{\hbar}+\nu_{p,r} \right),
\end{equation}
where
\begin{equation} \label{y}
  Y(\tau) = \int_{-\infty}^\infty d\eps \frac{\cos\left(\tau \eps/\hbar
  \right)}{\sqrt{\eps^2+\Db^2}}= 2 K_0\left(\tau/\tD \right).
\end{equation}
This equation, where all classical quantities involved are evaluated at
Fermi energy, contains detailed information about the variations
of the pairing gap. Note that $\Dt$ only depends on $G$ and $L$ through $\Db$.
$K_0(x)$ is the modified Bessel function of second kind. Through it, a new
characteristic ``pairing time'' associated with the pairing gap is introduced,
\begin{equation}
\tD = \frac{h}{2\pi\Db}.
\end{equation}
Since $K_0 (x) \propto \exp(-x)/\sqrt{x}$ for $x \gg 1$, the
Bessel function exponentially suppresses all contributions for
times $\tau \gg \tD$ (making the sum convergent).
The average part of the gap $\Db$ is thus playing, in this respect, a role very
similar to the temperature in the general theory of mesoscopic
fluctuations (cf Ref.~\cite{Monastra}). In contrast, $K_0 (x) \propto -\log (x)$
for $x \ll 1$, and short orbits (compared to $\tD$) are logarithmically enhanced.

Since the value of the actions depend on the shape of the mean--field potential,
Eq.~(\ref{dosc}) predicts, generically, fluctuations of the pairing gap
as one varies, for instance, the particle number, or the shape of the system
at fixed particle number. The fluctuations result from the interference between
the different oscillatory terms that contribute to $\Dt$. The symmetries of the
potential and the nature (integrable or chaotic) of the underlying classical motion
are crucial to understand the interference pattern. When the motion is regular
(integrable), continuous families of
periodic orbits having the same action, amplitude, etc, exist. The coherent
contribution to the sum (\ref{dosc}) of these families of
periodic orbits produces large
fluctuations. In contrast, in the absence of regularity or symmetries,
incoherent contributions of smaller amplitude coming from isolated unstable orbits
are expected. Moreover, aside the dependence on the regular or chaotic nature of
the single-particle motion, the presence or absence of universality in the statistical
properties of the fluctuations will depend on the dominance of short or long periodic orbits.

We will make below an analysis of the predictions of
Eq.~(\ref{dosc}) in the nuclear case, as the neutron number is
varied. Before, and in order to avoid at this stage a detailed
study of a particular system, we concentrate on a global analysis,
namely the typical size or root mean square (RMS) of the BCS gap
fluctuations in a generic mesoscopic system. The second moment of
the fluctuations may be expressed from Eq.~(\ref{dosc}) as
\begin{equation} \label{var}
  \secmom{\Dt} = \frac{\Db^2}{2\tH^2} \int_0^\infty d\tau Y^2(\tau)
  K(\tau),
\end{equation}
where $\tH = h/\delta$ is Heisenberg time ($\delta = \rb^{-1}$ is
the single--particle mean level spacing at Fermi energy), and
$K(\tau)$ is the spectral form factor, i.e. the Fourier transform
of the two-point density--density correlation function
\cite{Berry}.

The structure of the form
factor $K(\tau)$ is characterized by two different time scales.
The first one, the smallest of the system, is the period $\tm$ of
the shortest periodic orbit. %(roughly estimated in ballistic
%systems as twice the time for a classical particle moving at Fermi
%velocity to cross the system).
The form factor is zero for $\tau
\leq \tm$, and displays non-universal (system dependent) features
at times $\tm \leqa \tau \ll \tH$. As $\tau$ further increases, the
function becomes universal, depending only on the regular or
chaotic nature of the dynamics, and finally tends to $\tH$ when
$\tau \gg \tH$. The result of the integral (\ref{var}) thus
depends on the nature of the dynamics, and on the relative value
of $\tD$ with respect to $\tm$ and $\tH$. According to Anderson
criterion \cite{Anderson}, superconductivity exists if $\Db >
\delta$ (we are not interested here in the ultrasmall regime $\Db
< \delta$ \cite{ML} where the BCS theory fails). Then, $\Db >
\delta$ implies $\tD/\tH = \delta/2\pi\Db \ll 1$. Because the
Bessel function $K_0$ exponentially suppresses the amplitude for times
$\tau \gg \tD$, one can safely ignore the structure of the form
factor for times of the order or bigger than $\tH$, and use the so
called diagonal approximation of $K(\tau)$ \cite{Berry}. In the
simplest approximation, all the non--universal system--specific
features are taken into account only through $\tm$ \cite{Monastra},
and one can write $K(\tau) = 0$ for $\tau < \tm$ and, for $\tau \geq \tm$,
$K(\tau) = \tH$ for integrable systems and $K(\tau) = 2\tau$ for
chaotic ones with time reversal symmetry.

\begin{figure}[t]
\includegraphics[width=7.5cm,clip=true]{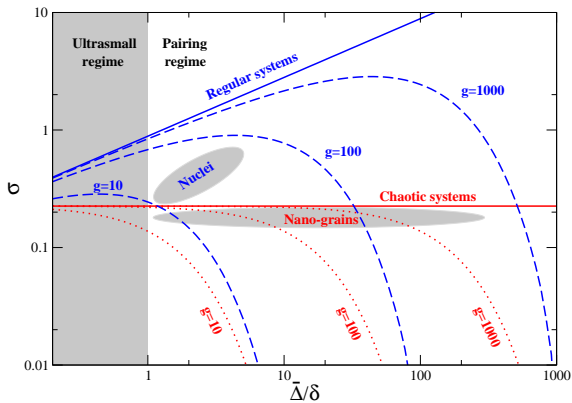}
\vspace{-0.3cm}\caption{Fluctuations of the pairing gap as a function of the mean
  value for mesoscopic systems (log-log scale; all quantities normalized with the
  single-particle mean spacing). The dashed (regular, in blue) and dotted
  (chaotic, in red) curves
  correspond to different values of the dimensionless conductance,
  $g$, and the limiting case of $g \rightarrow \infty$ is shown by
  solid lines. The results are valid in the pairing regime
  $\bar{\Delta}/\delta > 1$. Applications to Nuclei and
  Nano-grains are marked out. Ultracold atomic gases may be controlled to
  appear in major parts of the figure.} \label{Figgeneral}
\end{figure}

This finally gives the expressions for fluctuations of the pairing
gap (normalized to the single--particle mean level spacing),
$\sigma=\rms{\Dt}/\delta$, assuming regular dynamics \cite{comment_HO},
\begin{equation}
  \sigma_{\rm reg}^2 = \frac{\pi}{4} \frac{\Db}{\delta} F_0 \left( D \right) \ ,
  \label{secmomreg}
\end{equation}
and assuming chaotic dynamics,
\begin{equation}
  \sigma_{\rm ch}^2 = \frac{1}{2\pi^2} F_1 \left( D \right) \ ,
    \label{secmomch}
\end{equation}
where $F_n (D) = 1-\int_0^{D} x^n K_0^2(x) dx/\int_0^{\infty} x^n
K_0^2(x) dx$. The argument
\begin{equation}
  D = \frac{\tm}{\tD} = \frac{2 \pi}{g}
  \frac{\Db}{\delta}
  \label{D}
\end{equation}
is a system dependent quantity inversely proportional to the dimensionless
conductance, $g=\tH/\tm$, an intrinsic characteristic of the
system independent of the pairing coupling. $D$ can also be viewed as the 
system size divided by the coherence length of the Cooper pair, 
$\xi_0 = \hbar v_{\scriptscriptstyle F}/(2\Db)$, where $v_{\scriptscriptstyle F}$
is Fermi velocity. Equations
(\ref{secmomreg}) and (\ref{secmomch}), which together with Eq.~(\ref{dosc})
are the main results
of this study, show that the variance of the pairing gap is a
function of its normalized mean part, $\Db/\delta$, and of the
dimensionless conductance, $g$, as shown in
Fig.~\ref{Figgeneral}.
%This provides a generic description of
%mesoscopic fluctuations of the pairing gap in the weak coupling limit.

The monotonic function $F_n (D)$ has the following limiting
behaviors, $F_n (D)\rightarrow 1$ as $D\rightarrow 0$, whereas
$F_n (D) \rightarrow 0$ exponentially fast for $D \gg 1$. Thus, in
a system characterized by large $g$-value, $D \rightarrow 0$. In
this case $F_n (D) \rightarrow 1$, all dynamical system specific
properties disappear, and we obtain a "universal" (system
independent) behavior of the gap fluctuations given by the
prefactors in Eqs.~(\ref{secmomreg}) and (\ref{secmomch}), that
correspond to a pure uncorrelated Poisson sequence and to a GOE
random matrix spectrum, respectively (the latter, $\sigma_{\rm
ch}^2 = 1/(2 \pi^2)$, was obtained previously in Ref.~\cite{ML}).
This situation is shown by the solid lines in
Fig.~\ref{Figgeneral}: purely GOE fluctuations imply a constant
amplitude of the normalized fluctuations of the pairing gap, whereas an
increase with $\Db /\delta$ is seen for systems with uncorrelated spectra.
In contrast, in the generic case of systems characterized by finite values of $g$,
$F_n$, and therefore the pairing fluctuations, may significantly
deviate from universality (cf Fig.~\ref{Figgeneral}). Thus, in
general, pure statistical models (like GOE) do not provide an
adequate description of the pairing fluctuations.

We shall now apply these results to different physical situations.
Our first example is a system dominated by regular dynamics,
namely ground states of atomic nuclei, which bring
the best experimental data available at present on the
superfluidity of finite Fermi systems. The ground-state
superfluidity of atomic nuclei implies a mass difference between
systems with an even and odd number of particles. The connection between the
pairing gap and the mass differences is given by the
three--point measure $\Delta_3 (M) = B(M) - [B(M+1) + B(M-1)]/2$,
where $M$ is the odd neutron $N$ or proton $Z$ number. In the presence of other
possible interactions, this quantity has been
shown to be a very good measure of pairing correlations \cite{doba}.
$\Delta_3$ is shown in Fig.~\ref{Fignucl} for neutrons.
The average dependence of the neutron and proton gaps is well approximated,
from experimental data, by
\begin{equation}
  \Db = \frac{2.7}{A^{1/4}}~{\rm MeV} \ ,
  \label{Dbeq}
\end{equation}
where $A=N+Z$ is the total number of nucleons \cite{comment_Delta}.
We notice a rather strong variation around the
average value. The $A$ dependence of the RMS of the experimental
pairing fluctuations is shown in the inset of Fig.~\ref{Fignucl}.

In order to evaluate the RMS of the pairing fluctuations from the
theoretical expressions, Eqs.~(\ref{secmomreg}) and
(\ref{secmomch}), we need the following estimates of nuclear
properties (for one nucleon type): mean level spacing, $\delta
\approx 50/A~{\rm MeV}$, and dimensionless conductance $g \approx
1.6 A^{2/3}$ \cite{masses}. This gives $D \approx 0.21 A^{1/12}$,
which ranges from $0.27$ for $A=25$ to $0.33$ for $A=250$. Though
these values clearly set atomic nuclei in the regime $\tm/\tD<1$
where pairing fluctuations are important, we are still far from
$D=\tm/\tD=0$, so that significant deviations from universality are expected.
By inserting these estimates in Eqs.~(\ref{secmomreg}) and
(\ref{secmomch}) the fluctuations are easily evaluated, assuming
regular or chaotic dynamics. The resulting curves are compared to
the experimental one in the inset of Fig.~\ref{Fignucl}. Note,
as expected \cite{comment_Mixed}, the good agreement between the regular dynamics
and the experimental curve, either in the overall amplitude as well
as in the $A$--dependence.

One may go beyond a statistical description, and use Eq.~(\ref{dosc})
to obtain a detailed description of the fluctuations. For that purpose,
we assume for the nuclear mean field a simple hard-wall cavity potential.
The shape of the cavity at a given number of nucleons
is fixed by minimization of the energy against quadrupole, octupole
and hexadecapole deformations. To simplify, we take $N=Z$. The periodic
orbits of a spherical cavity are used in Eq.~(\ref{dosc}),
with modulations factors that take into account deformations and
inelastic scattering \cite{creagh}. We set the average of $\Dt$ to zero,
as was done with the experimental data (although of interest by
itself, we will not consider its behavior here).
In Fig.~\ref{Fignucl} we compare the theoretical result $\Dt (N)$ to
the experimental value averaged over the different isotopes at a given $N$.
The agreement is excellent; the theory describes all the main features
observed in the experimental curve.

Our second example are the superconducting properties of
nano-sized clean (ballistic) metallic grains \cite{diffusive},
where we may expect the dynamics to be chaotic, see
Ref.~\cite{NanoRev}.
%Experiments in the 90's have explored the
%superconducting properties of nanometer scale aluminum grains
%\cite{RBT}. The existence of a superconducting gap was
%demonstrated in the regime $\Db > \delta$, whereas no gap was
%observed when $\Db < \delta$ (the transition occurs around $N \sim
%5000$, where $N$ is the number of conduction electrons in the
%grain).
The existence of a superconducting gap was
demonstrated in the regime $\Db > \delta$ \cite{RBT}, whereas no gap was
observed when $\Db < \delta$ (the transition occurs around $N \sim
5000$, where $N$ is the number of conduction electrons in the
grain).
The $N$ dependence of the average gap $\Db$ is poorly
understood. We will adopt for grains the thin-film value $\Db
\approx 0.38 \times 10^{-3}$~eV \cite{NanoRev}. The mean level
spacing is $\delta = (2\ef)/(3N) \approx 2.1/N$~eV, whereas $g
\approx 2.6 N^{2/3}$. Eq.~(\ref{D}) gives $D \approx
4.4 \times 10^{-4} N^{1/3}$, which ranges from $0.05$ to $0.02$
when $N$ varies between $10^{3}$ and $10^{5}$. This means that the
variance will be close to its ``universal'' value obtained by
setting $F_1 (D) = 1$ in Eq.~(\ref{secmomch}), namely
$\sigma^2_{\rm ch} = 1/2 \pi^2$. Their typical range of variation
is represented in Fig.~\ref{Figgeneral}.

In the case of chaotic dynamics, we don't have explicit
experimental or numerical data to compare with. We can,
alternatively, compute the fluctuations of the condensation
energy, defined as the total energy difference between the paired
and unpaired system. In the universal chaotic limit, our results
are in good agreement with the numerical calculations of Sierra
{\it et~al.} \cite{Dukelsky}, where they use Richardson's
solution of the pairing problem and random matrix theory (GOE) for
generating the single--particle spectrum.

Superfluidity in ultracold atomic gases is currently intensively studied,
and provides our third example. The confinement potential can be
externally controlled to create regular as well as chaotic
dynamics,
and the atom-atom interaction, $a$, can be tuned around
the Feshbach resonance. Since both particle number and
interaction strength are experimentally controlled parameters, the fluctuations
may appear in major
parts of Fig.~\ref{Figgeneral}. We estimate $\delta = (2\ef)/(3N)$ and
$g=\frac{1}{2}(3N)^{2/3}$; in the dilute BCS region
$\Db=(2/e)^{7/3} \ef \exp\left( -\pi/2 \kf |a| \right)$ \cite{Gorkov},
with $\kf$ the Fermi wavevector, giving
$D=2\pi (2/e)^{7/3} (3N)^{1/3} \exp \left(-\pi/2\kf |a| \right)$.
Recent experiments using Li$^6$ reach $\kf |a| = 0.8$ \cite{KetterleLi6},
implying negligible fluctuations for typical values of $N
\sim 10^6$. %Reducing $\kf |a|$ by a factor of 4 to
%$\kf |a|=0.2$ and the particle number to
%$N = 10^3$ yields for generic regular systems fluctuations that are
%on the same magnitude as the mean pairing gap, $\sigma_{\rm reg} \approx \Db /
%\delta$ \cite{comment_HO}.
Reducing to $\kf |a|=0.2$ and $N = 10^4$ yields for generic regular systems
fluctuations that are 
on the same magnitude as the mean pairing gap, $\sigma_{\rm reg} \approx 0.5
\Db / \delta$ \cite{comment_HO}.

To conclude, we have presented an explicit semiclassical theory for the pairing
gap fluctuations. These are generically dominated by system specific features not
included in purely statistical models.
Different possible regimes, as well as the influence of order/chaos dynamics, were
investigated, in particular for the typical size of the fluctuations
(Fig.~\ref{Figgeneral}). The present theory provides, for the first time,
analytic predictions, valid for a wide range of physical situations; it also
compares very favorably with available experimental data.

P.L. acknowledges support by grants ANR--05--Nano--008--02,
ANR--NT05--2--42103 and by the IFRAF Institute.


\vspace{-0.8cm}
\begin{thebibliography}{99}
\vspace{-0.6cm}
%------------------------------
\bibitem{BCS}
  J. Bardeen, L.N. Cooper and J.R. Schrieffer, Phys. Rev. {\bf 106}, 162
  (1957); \textit{ibid} {\bf 108}, 1175 (1957).
%------------------------------
\bibitem{Pines}
  A. Bohr, B.R. Mottelson and D. Pines, Phys. Rev. {\bf 110}, 936 (1958).
%------------------------------
\bibitem{RevNucl}
  D. Dean and M. Hjort-Jensen, Rev. Mod. Phys. {\bf 75}, 607 (2003).
%------------------------------
\bibitem{NanoRev}
  J. von Delft and D.C. Ralph, Phys. Rep. {\bf 345}, 61 (2001).
%------------------------------
\bibitem{RevFermiGas}
  C.A Regal, M. Greiner and D.S. Jin, Phys. Rev. Lett. {\bf 92}, 040403
  (2004); M.W. Zwierlein {\it et al.}, Phys. Rev. Lett. {\bf 92}, 120403
  (2004); H. Heiselberg and B.R. Mottelson, Phys. Rev. Lett. {\bf 88}, 190401
  (2002); G.M. Bruun and H. Heiselberg, Phys. Rev. A {\bf 65}, 053407 (2002).
%------------------------------
\bibitem{Brink}
  D. Brink and  R.A. Broglia, \textit{Nuclear Superfluidity: Pairing in
  Finite Systems} (Cambridge Univ. Press, 2005).
%------------------------------
\bibitem{Anderson}
  P.W. Anderson, J. Phys. Chem. Solids B {\bf 11}, 26 (1959).
%------------------------------
\bibitem{RBT}
  D.C. Ralph, C.T. Black and M. Tinkham, Phys. Rev. Lett. {\bf 74}, 3241
  (1995); \textit{ibid} {\bf 76}, 688 (1996).
%------------------------------
%\bibitem{shell}
%------------------------------
%\bibitem{Zwierlein}
%  M.W. Zwierlein {\it et al.}, Nature {\bf 442}, 54 (2006).
%------------------------------
\bibitem{semicl}
  M. Brack and R.K. Bhaduri, \textit{Semiclassical Physics} (Addison and
  Wesley, Reading, 1997).%; K. Richter, \textit{Semiclassical Theory of
%    Mesoscopic Quantum Systems} (Springer, Berlin, 2000).
%------------------------------
\bibitem{PapenBulgacYu}
 C. A. R. S\'a de Melo, M. Randeria, and J. R. Engelbrecht, 
 Phys. Rev. Lett. {\bf 71}, 3202 (1993);
 A. Bulgac and Y. Yu, Phys. Rev. Lett. {\bf 88}, 042504 (2002).
%------------------------------
\bibitem{Monastra}
  P. Leboeuf and A.G. Monastra, Ann. Phys. {\bf 297}, 127 (2002).
%------------------------------
\bibitem{Berry}
  M.V. Berry, Proc. Roy. Soc. Lond. {\bf A 400}, 229 (1985).
%------------------------------
\bibitem{ML}
  K.A. Matveev and A.I. Larkin, Phys. Rev. Lett. {\bf 78}, 3749 (1997).
%------------------------------
\bibitem{comment_HO}
We assume a generic regular system; the analysis does not apply
to the harmonic oscillator, whose form factor is pathological.
%------------------------------
\bibitem{doba}
  J. Dobaczewski {\it et al.}, Phys. Rev. C {\bf 63}, 024308 (2001).
%------------------------------
\bibitem{comment_Delta}
The traditionally employed expression $ \bar{\Delta} =12/\sqrt{A}$
MeV (obtained if also even particle number in the three-point
formula are included) gives similar fluctuations.
%------------------------------
\bibitem{masses}
  O. Bohigas and P. Leboeuf, Phys. Rev. Lett. {\bf 88}, 092502 (2002).
%------------------------------
\bibitem{comment_Mixed}
Our theoretical description is not accurate enough to conclude
about the possible presence and influence in the pairing gap of small
chaotic components, as in Ref.\cite{masses} for the nuclear masses.
%------------------------------
\bibitem{creagh}
  S. C. Creagh, Ann. Phys. (N.Y.) {\bf 248}, 60 (1996); P. Leboeuf,
  Lect. Notes Phys. {\bf 652}, Springer, Berlin Heidelberg 2005, p.245,
  J. M. Arias and M. Lozano (Eds.).
%------------------------------
\bibitem{diffusive}
%Diffusive nano--grains may be treated as well within our formalism
%by an appropriate change of the spectral form factor, $K(\tau)$.
Diffusive nano--grains may be treated in our formalism
by an appropriate change of the spectral form factor, $K(\tau)$.
%------------------------------
\bibitem{Dukelsky}
  G. Sierra {\it et al.}, Phys. Rev. B
  {\bf 61}, 11890 (2000).
%------------------------------
\bibitem{Gorkov}
  L.P. Gor'kov and T.K. Melik-Barkhudarov, Sov. Phys. JETP {\bf 13}, 1018 (1961).
%------------------------------
\bibitem{KetterleLi6}
  C.H. Schunck {\it et al.}, Phys. Rev. Lett. {\bf 98}, 050404 (2007).
%------------------------------
\bibitem{NuclMass}
  G. Audi, A.H. Wapstra and C. Thibault, Nucl. Phys. {\bf A729}, 337 (2003).
%------------------------------

\end{thebibliography}
\end{document}